\documentclass[11pt]{article}

\usepackage[english]{babel}
\usepackage{graphicx}
\usepackage{pstricks}

\begin{document}
\hfill{FTUV-01-0915}
    
\hfill{IFIC-01-51} 

\vspace{1cm}

\begin{center}
{\Large\bf Single Spin Asymmetry Parameter from Deeply Virtual Compton
Scattering of Hadrons up to Twist-3 accuracy:\newline
 I. Pion case}

\vspace{.5cm}
\rm{
I.V.~Anikin$^{a,\,b}$ , D.~Binosi$^{a,\ c}$ ,
R.~Medrano$^{a}$, S.~Noguera$^{a}$,
V.~Vento$^{a,\,c,\,d}$}\\
\vspace{.5cm}
{\it $^{a}$Departamento de F\'{\i}sica Te\'{o}rica,
Universidad de Valencia, \\
E-46100 Burjassot, Spain}\\
{\it $^{b}$ Bogoliubov Laboratory of Theoretical Physics,\\
Joint Institute for Nuclear Research,
141980 Dubna, Russia }\\
{\it $^{c}$ IFIC-CSIC, E-46071 Valencia, Spain}\\
{\it $^{d}$ School of Physics, Korea Institute for Advanced Study,\\ 
Seoul 130-012, Korea}
\end{center}

\begin{abstract}

The study of  Deeply Virtual Compton Scattering has shown that electromagnetic
gauge invariance requires, to leading order, not only twist two but additional
twist three contributions. We apply this analysis and,  using the
Ellis-Furmanski-Petronzio factorization scheme, compute the single (electron)
spin asymmetry arising in the collision of  longitudinally polarized electrons
with hadrons up to twist 3 accuracy. In order to simplify the kinematics
we restrict the actual calculation to pions in the chiral limit.  The
process is described in terms of the generalized parton  distribution functions
which we obtain within a bag model framework.

\end{abstract}

\section{Introduction.}

Hard reactions  provide important information for unveiling the structure of
hadrons. The large virtuality, $Q^2$, involved in these processes allows the
factorization of the hard (perturbative) and soft (nonperturbative)
contributions in their amplitudes. Therefore these  reactions are receiving
great attention by the hadronic physics  community. Among the hard processes
one, which merits to be singled out, is the Deeply Virtual Compton  Scattering
(DVCS) because it can be expressed, in the asymptotic regime, in  terms of the
so called Generalized Parton Distributions (GPDs)  \cite{Mul94,Ji97b,Rad97}.
The GPDs describe non-forward matrix elements of light-cone operators and
therefore measure the response of the internal structure of the hadrons to the
probes. Moreover DVSC is instrumental in the experimental  interpretation of
the angular momentum  sum rule \cite{Ji97a}.

It has been shown  that the implementation of gauge invariance in the analysis
of the DVCS amplitude in the asymptotic regime, {\it i.e.}
 large  virtuality of the
incoming  photon, requires the inclusion of twist-3  contributions 
\cite{Gui98}-\cite{Kiv01b}. Let us explain the reason in a brief manner. 
In leading twist, in the Bjorken limit, the Lorentz structure of the hard subgraph
of the DVCS amplitude's leading diagram has, at large $Q^2$, the form of a
transverse projector. The virtual photon momentum in the form of the Sudakov
decomposition  contains a transverse component too. Thus their contraction does
not vanish and electromagnetic gauge invariance is violated. To restore it, 
next-to-leading  order terms in the asymptotic expansion to the DVCS amplitude,
which are proportional to the transverse component of the momentum transfer,
have to be included. These terms, which are twist-3, give rise to the dominant
contribution in some observables. One of them is the single spin asymmetry
(SSA), which arises in the collision of longitudinally polarized electrons with
hadrons, and which we will analyze in here \footnote{Recently,  a detailed 
report with the description of 
all possible asymmetries has appeared in the literature 
(see Ref.\cite{Goe01}).}.

In this first paper we deal with a spin zero massless target, the pion.
The same procedure could be applied to the scattering off unpolarized nucleons.
However this calculation would require keeping the mass terms of the nucleon,
a complication which we want to avoid at present \cite{Ani01}. Moreover, in the
case of polarized nucleons, one could study besides the SSA other asymmetries,
however the complexity of the analysis, with the existence of many different
GPDs, is postponed for a future publication \cite{Ani01}. In order to calculate
the  twist-2 and  twist-3\footnote{We stress that throughout this paper we 
mean twist-3, which include both the kinematical and dynamical 
twist-3 contributions.} 
GPDs  contributing to the SSA, a crucial ingredient
of the calculation,  we  use  an MIT bag model scheme with boosted wave
functions.

The plan of our paper is as follows. In section 2 we describe the kinematics
and introduce the appropriate notations. The DVCS amplitude for the pion,
including up to  twist-3 contributions, is presented in section 3. We follow
the description of Ref.\cite{Ani00}, where, using a generalization of the
Ellis-Furmanski-Petronzio (EFP) factorization scheme \cite{Ell82}, the
complete gauge invariant DVCS amplitude has been obtained. In  section 4 we
outline  the basic ingredients of our approach and calculate all the
parametrizing  functions (GPDs). Then in section 5 we give the numerical
estimates of  the  SSA parameter for  the case under study and discuss our
results.

\section{$e h\to\gamma e h $ process: kinematics and notations.}

Our starting point is electron-hadron scattering into real photon,
electron and hadron,
\begin{eqnarray}
\label{reaction}
e(k)+ \pi(p) \to\gamma(q^{\prime})+ e(k^{\prime})+\pi(p^{\prime}),
\end{eqnarray}
assuming that the electron is longitudinally polarized.
We will consider the electron to be massless,
\begin{eqnarray}
k^2=k^{\prime\, 2}=0,
\end{eqnarray}
and, that the hadron is a pion, which we take also to be massless
(chiral limit),
\begin{eqnarray}
p^2=p^{\prime\, 2}=m_{\pi}^2\to 0.
\end{eqnarray}
In  QED, to lowest order, the reaction (\ref{reaction}) takes place via
the Bethe-Heitler process, where the final real photon is emitted by
one of the electrons, and the virtual Compton process
\begin{eqnarray}
\label{dvcs}
\gamma^*(q)+\pi(p) \to\gamma(q^{\prime})+\pi(p^{\prime}),
\end{eqnarray}
which becomes the DVCS process if the square of virtual photon momentum
$q=k-k^{\prime}$ is very large, {\it i.e.}
\begin{eqnarray}
Q^2=-q^2\to\infty .
\end{eqnarray}

For the Bethe-Heitler process the amplitude is given by
\begin{eqnarray}
\label{BH}
T_{(BH)}=-e^3
\bar u(k^{\prime}) \biggl\{
\hat\varepsilon^*\frac{\hat k -\hat\Delta}{(k-\Delta)^2}
\gamma_{\mu}+
\gamma_{\mu}\frac{\hat k^{\prime}+\hat\Delta}{(k^{\prime}+\Delta)^2}
\hat\varepsilon^*
\biggr\} u(k)
\frac{1}{\Delta^2} \Gamma_{\mu}(p,p^{\prime}),
\end{eqnarray}
where
\begin{eqnarray}
\label{Delta}
\Delta = p^{\prime} - p,
\end{eqnarray}
and the electromagnetic vertex of the pions takes the standard form
\begin{eqnarray}
\label{emv}
\Gamma_{\mu}(p,p^{\prime})=(p+p^{\prime})_{\mu}F_+(\Delta^2).
\end{eqnarray}
$F_+(\Delta^2)$ is the electromagnetic form factor of the pion.
The Virtual Compton process amplitude is given by
\begin{eqnarray}
\label{VC}
T_{(VC)}=
\bar u(k^{\prime})\gamma_{\mu} u(k)
\frac{1}{Q^2} T_{\mu\alpha}^{DVCS}\varepsilon_{\alpha}^* ,
\end{eqnarray}
where $T_{\mu\nu}^{DVCS}$ corresponds to the DVCS subprocess.

To describe the reaction (\ref{reaction})
it is useful to introduce the following dimensionless fractions
\begin{eqnarray}
&&x=\frac{Q^2}{2p\cdot q},  \quad y=\frac{p\cdot q}{p\cdot k },
\nonumber\\
&&z=\frac{p\cdot p^{\prime}}{p\cdot q} \,\,\ or \,\,\,
\frac{p\cdot q^{\prime}}{p\cdot q} .
\end{eqnarray}
These fractions can be related with the Mandelstam variables of
reaction (\ref{reaction}) and of the DVCS subprocess.
Indeed, if we introduce the following variables
\begin{eqnarray}
\hat s= (q+p)^2, \quad \hat t=(p^{\prime}-p)^2,
\end{eqnarray}
for the DVCS subprocess, and
\begin{eqnarray}
S=(k+p)^2\approx 2k\cdot p,
\end{eqnarray}
for the reaction (\ref{reaction}), the following relations hold
\begin{eqnarray}
Q^2=xyS, \quad \hat s=(1-x)yS, \quad \hat t=ySz.
\end{eqnarray}

Since we neglect the pion mass, the most suitable system of reference is the
center of mass system, where
\begin{eqnarray}
&&k=E(1,\sin\beta,0,\cos\beta), \quad q= (Q_0,0,0,-E_p),
\nonumber\\
&&p=E_p (1,0,0,1 ),
\quad
p^{\prime}=E_{p^{\prime}}
(1,\sin\theta\cos\phi,\sin\theta\sin\phi,\cos\theta ),
\end{eqnarray}
and
\begin{eqnarray}
&& E_p=\frac{\hat s + Q^2}{2\sqrt{\hat s}}, \quad
E_{p^{\prime}}=E_{q^{\prime}}=\frac{1}{2}\sqrt{\hat s}, \\
\nonumber\\
&&E=\frac{S - Q^2}{2\sqrt{\hat s}}, \quad
Q_0=\frac{\hat s - Q^2}{2\sqrt{\hat s}}.
\end{eqnarray}
Moreover, we define $\phi$ to be the angle between the plane formed by
the three-dimensional vectors $q^{\prime}$ and $k$ (leptonic plane)
and the plane formed by the three-dimensional vectors $q^{\prime}$ and
$p^{\prime}$ (hadronic plane).
It can then be easily seen  that the other angles satisfy the following relations
\begin{eqnarray}
\cos\theta=1-2z, \quad \cos\beta=1- \frac{2(1-x)}{1-xy}.
\end{eqnarray}

\section{DVCS amplitude off pions up to twist-3.}

In this section we focus on the DVCS amplitude off pions.
We start from the expression for the virtual Compton scattering amplitude
which can be written as usually in the form 
\begin{eqnarray}
\label{comm_amp}
T_{\mu\nu}=i\int d^4z \,e^{iq^{\prime}z} \, 
\langle p^{\prime} | T J_{\mu}(0)J_{\nu}(z) |p \rangle ,
\end{eqnarray}
where $J_{\mu}$ is the electromagnetic quark current:
\begin{eqnarray}
\label{curr}
J_{\mu}(x)=\bar\psi(x)\,{\cal Q}\, \gamma_{\mu}\, \psi(x).
\end{eqnarray} 
In Eq.(\ref{curr}) ${\cal Q}$ is the charge quark matrix, which 
is equal to 
\begin{eqnarray}
\frac{1}{6}(1 + 3\tau_3) \quad {\rm for}\,\, SU_F(2)
\end{eqnarray} 
and to 
\begin{eqnarray}
\frac{1}{2}(\lambda_3 +\frac{1}{\sqrt{3}}\lambda_8)\quad 
{\rm for}\,\, SU_F(3).
\end{eqnarray}
Here $\tau_i$ and $\lambda_i$ are the conventional Pauli and
Gell-Mann matrices for two and three flavors respectively.

As discussed before  a gauge invariant DVCS amplitude cannot be written down
unless the twist three contributions to the amplitude are  taken into account.
Here we would like to recall, briefly, the results obtained
in Ref.\cite{Ani00} and reproduced by several groups 
\cite{Pen00, Bel00, Rad00}.

The main point of these analyses is that, the violation of the photon gauge
invariance, is proportional to the  non-zero transverse component of the
virtual  photon momentum \cite{Gui98}. In other words, the convolution of the
leading order DVCS amplitude with the virtual photon momentum is  proportional
to the first degree of transversity that corresponds to  twist-3  \footnote{In
the present case, the transversity is the transverse component  of momentum
transfer.}. Hence to obtain the complete gauge invariant DVCS amplitude  we
have to consider all terms which are linear combinations of the
transversity. This situation is not surprising and  similarity with the
transverse polarization in deep inelastic scattering off nucleons can be
recalled \cite{Ans95, Ter96}.

\begin{figure}
\begin{center}
\includegraphics[width=12.0cm]{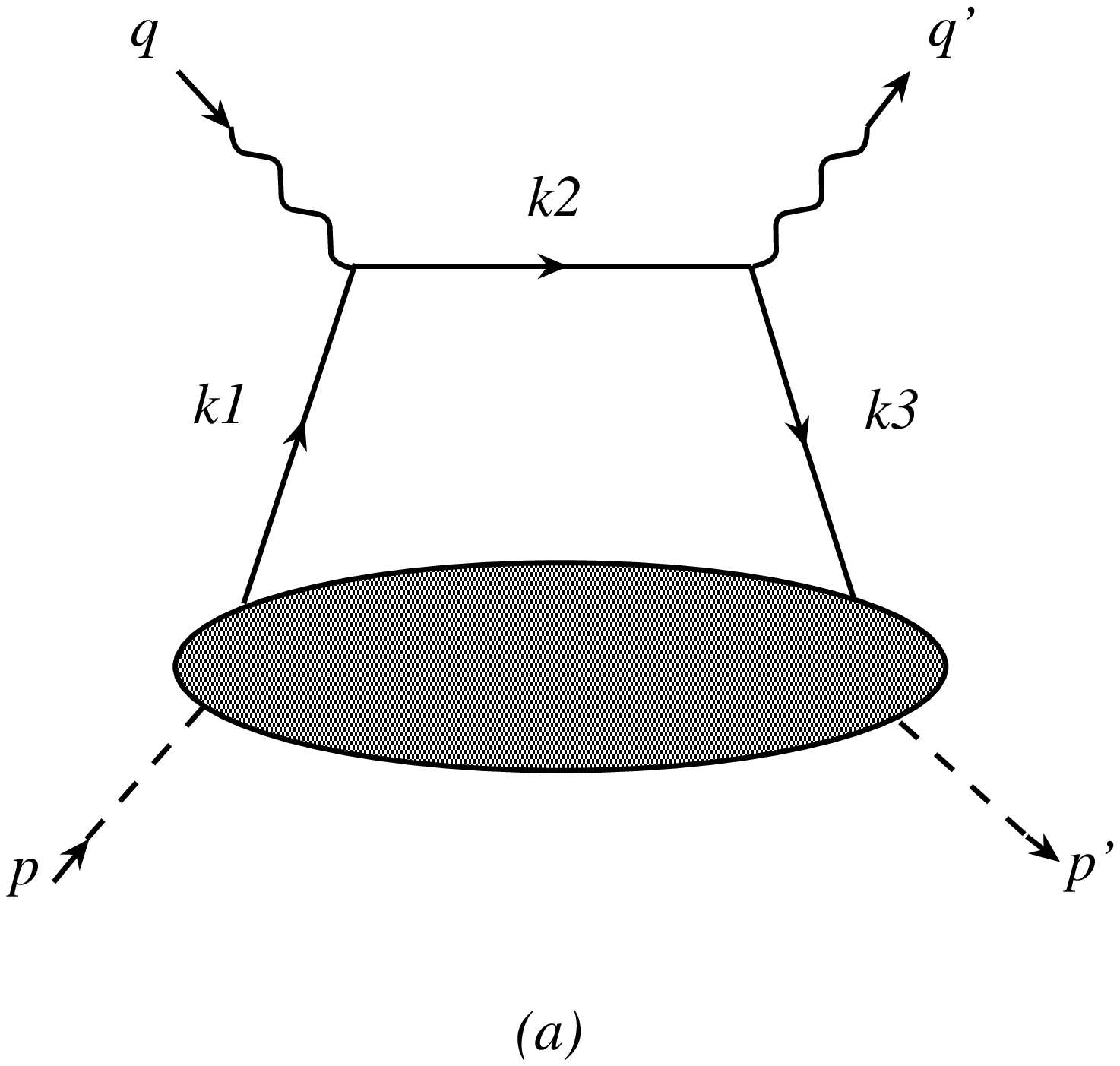}
\vspace{-1.5cm}
\includegraphics[width=12.0cm]{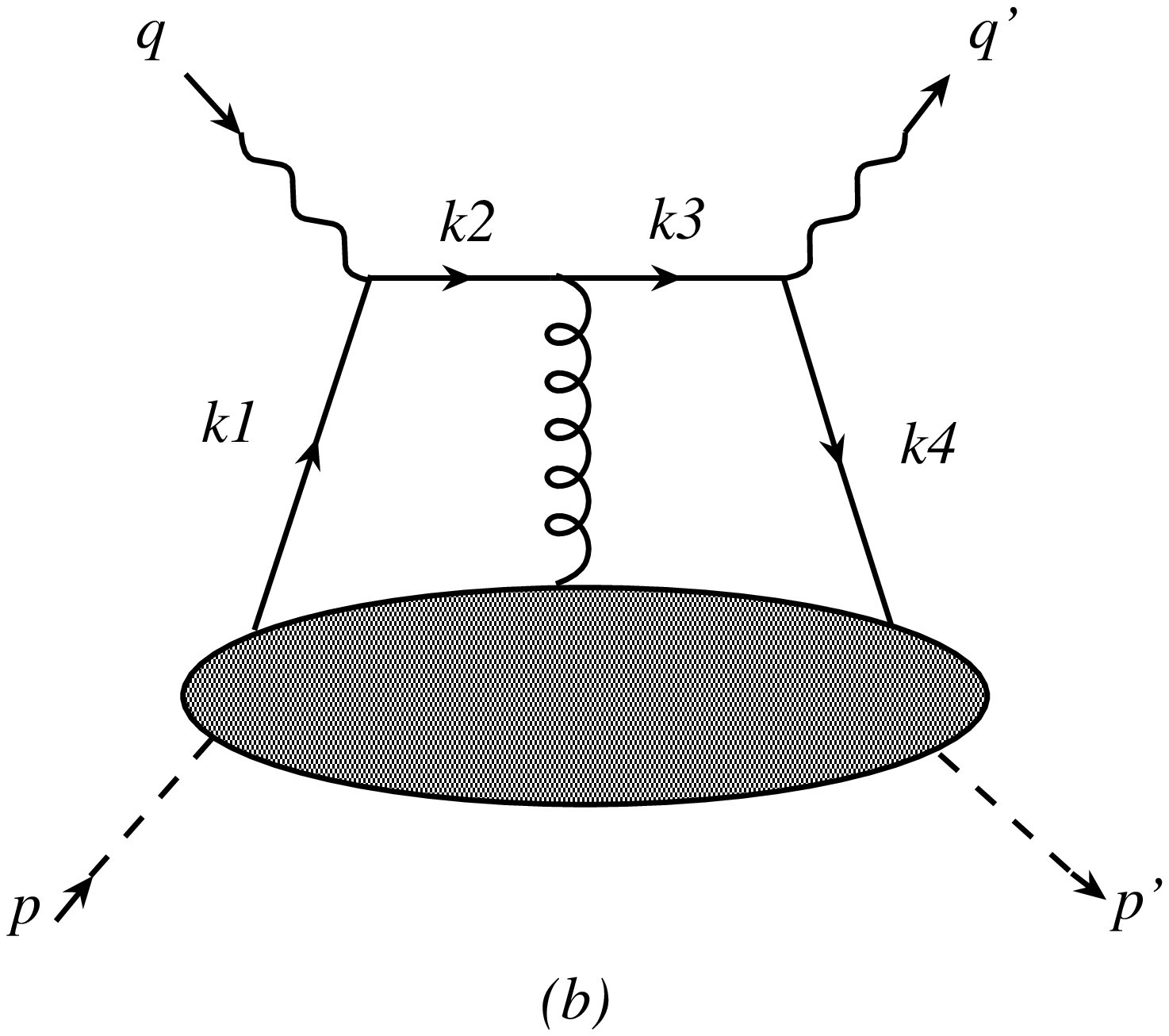}
\vspace{1.0cm}
\caption{The diagrams contributing to the DVCS amplitude in the EFP 
factorization scheme, $k1=xP-\Delta/2$, 
$k3=xP+\Delta/2$ for $(a)$ ;
$k1=x_1P-\Delta/2$, $k4=x_2P+\Delta/2$ for $(b)$ }
\label{fig1}
\end{center}
\end{figure}

Therefore, in order to preserve the 
electromagnetic gauge invariance up to leading 
order within the generalized EFP factorization scheme,  
one must add to diagram (a) of Fig.\ref{fig1},
consisting of a hard part with two quark legs
\footnote{In this paper we deal with the Born diagrams only.
The EFP factorizing scheme for the general case can be found in 
\cite{Ani00}},  the diagram consisting of a hard part
with two quark legs and one transverse gluon (see diagram (b) of
the same Figure). This latter  diagram is entirely  twist 3, while
diagram (a) contains, besides the standard twist-2 term produced
by the good components of the quark fields and collinear parton momenta, a
twist-3 term, which can be related to the quark gluon 
contribution of (b) by means of the equations of motion.

After performing the $T-$product for the two electromagnetic currents and 
going from the four-dimensional integration over $z$
to the one-dimensional integration over the $x-$fraction 
the amplitudes of diagrams (a) and (b) may be written as \cite{Ani00,Efr84}:
\begin{eqnarray}
\label{1.1}
&&T_{\mu\nu}^{(a)}+T_{\mu\nu}^{(b)}=
\int dx{\rm tr}\biggl\{ E_{\mu\nu}(x P) \Gamma (x) \biggr\}
+
\nonumber\\
&&\int dx_1 dx_2{\rm tr}\biggl\{
E_{\mu\rho\nu}(x_1 P, x_2 P) \omega_{\rho\rho^{\prime}}
\Gamma_{\rho^{\prime}} (x_1, x_2) \biggr\},
\end{eqnarray}
where $\omega_{\rho\rho^{\prime}}=\delta_{\rho\rho^{\prime}}-
n_{\rho^{\prime}} P_{\rho}$,
and
\begin{eqnarray}
\label{2.1}
&& E_{\mu\nu}(x P) = \gamma_{\mu} S(xP-\frac{\Delta}{2}+q) \gamma_{\nu}+
"crossed"
\nonumber\\
&&E_{\mu\rho\nu}(x_1 P,x_2 P) = 
\gamma_{\mu} S(x_1P-\frac{\Delta}{2}+q)
\gamma_{\rho} S(x_2P-\frac{\Delta}{2}+q)\gamma_{\nu}+
"crossed" 
\nonumber\\
&&\Gamma(x)=-\int d\lambda \,
e^{ i( x+\xi)\lambda }
\sum\limits_{a=1}^{N_f}\, e_a^2 \,
\langle p^{\prime} |
\psi_{a}(\lambda n) \bar\psi_{a}(0)| p \rangle ,
\nonumber\\
&&\Gamma^{\rho^{\prime}}(x_1, x_2)=
-\int d\lambda_1 d\lambda_2 \,
e^{ i( x_1+\xi) \lambda_1
+i(x_2-x_1) \lambda_2 }
\nonumber\\
&&\sum\limits_{a=1}^{N_f}\, e_a^2 \,\langle p^{\prime} |
\psi_{a}(\lambda_1 n)
\stackrel{ \leftrightarrow }
{ D^{\rho^{\prime}} }(\lambda_2 n)
\bar\psi_{a}(0)
| p \rangle.
\end{eqnarray}
Here $D_{\mu}$ is the QCD covariant derivative in the fundamental
representation
\footnote{We follow the notation of ref.\cite{Ani00}}.

The use of the QCD equations of motion lead to the following expectation values
\begin{eqnarray}
\langle p^{\prime} |
\overrightarrow{
\hat D(z)} \psi(z) \bar\psi(0) | p \rangle = 0 ,
\quad
\langle p^{\prime} | \psi(z) \bar\psi(0)
\overleftarrow{\hat D(0)} | p \rangle  = 0.
\end{eqnarray}
Keeping only the non-zero (axial and vector) projections
of the quark and quark-gluon correlators, we are able to express
the tree-body (quark-gluon) parametrizing functions in terms of
the two-body (quark) parametrizing functions.
As a result of this trick, the amplitude corresponding to
diagram (b), expressed by means of two-body parametrizing functions,
grouped together with  the amplitude corresponding to diagram (a),
expressed also by two-body parametrizing functions, leads to a
gauge invariant DVCS amplitude, which  reads
\begin{eqnarray}
\label{amp}
T_{\mu\nu}^{DVCS} =
-\frac{1}{2P\cdot Q}\int dx
\Biggl(
\frac{1}{x-\xi+i\epsilon} + \frac{1}{x+\xi-i\epsilon}
\Biggr)
{\cal T}_{\mu\nu},
\nonumber\\
\end{eqnarray}
where
\begin{eqnarray}
&&{\cal T}_{\mu\nu} =
H_1(x)
\Biggl(
-2\xi P_{\mu}P_{\nu}  -
P_{\mu}Q_{\nu} - P_{\nu}Q_{\mu} +
\Biggr.
\nonumber\\
\Biggl.
&&g_{\mu\nu}(P\cdot Q) - \frac{1}{2}P_{\mu}\Delta_{\nu}^{T} +
\frac{1}{2}P_{\nu}\Delta_{\mu}^{T}
\Biggr) -
\nonumber\\
&&H_3(x)
\Biggl(
\xi P_{\nu}\Delta_{\mu}^T + 3\xi P_{\mu}\Delta_{\nu}^T
+ \Delta_{\mu}^{T}Q_{\nu} + \Delta_{\nu}^{T}Q_{\mu}
\Biggr) -
\nonumber\\
&&\frac{\xi}{x}H_A(x)
\Biggl(
3\xi P_{\mu}\Delta_{\nu}^{T} -
\xi P_{\nu}\Delta_{\mu}^{T} -
\Delta_{\mu}^{T}Q_{\nu} + \Delta_{\nu}^{T}Q_{\mu}
\Biggr).
\nonumber
\end{eqnarray}
Here the parametrizing functions (GPDs) are,
\begin{eqnarray}
H_i(x)=\sum\limits_{a=1}^{N_f} e_a^2 H_i^a(x), \,\, i=\{1,3,A\} ,
\,\, N_f = {\rm number \,\, of \,\, flavors}, 
\end{eqnarray} 
which were defined in Ref.\cite{Ani00}, and whose calculation we show in 
Section 5.
The DVCS amplitude is complex, due to the factor in front
of ${\cal T}_{\mu\nu}$ in Eq.(\ref{amp}), leading therefore to  a
non vanishing SSA, which could be seen in the collision of longitudinally
polarized electron beams with pions.

\section{Single spin asymmetry parameter.}

We next calculate the SSA parameter \cite{Ji97b}
as a function of the angle $\phi$ between the leptonic and hadronic planes.
The SSA parameter is given by
\begin{eqnarray}
\label{ssa}
{\cal A}_L=\frac{d\sigma(\rightarrow)-d\sigma(\leftarrow)}
{d\sigma(\rightarrow)+d\sigma(\leftarrow)},
\end{eqnarray}
where $d\sigma(\rightarrow, \leftarrow)$ denotes the differential
cross section with different helicities for electrons.

The difference of cross sections in the
numerator of (\ref{ssa}) is defined by the imaginary part
of the convolution of the leptonic tensor with the hadronic tensor and
consists of two terms
\begin{eqnarray}
\label{difcross}
d\sigma(\rightarrow)-d\sigma(\leftarrow)=
\Delta d\sigma^{(I)}(\rightarrow;\leftarrow) +
\Delta d\sigma^{(S)}(\rightarrow;\leftarrow).
\end{eqnarray}
The first term in (\ref{difcross}),
$\Delta d\sigma^{(I)}(\rightarrow;\leftarrow)$,
 emanates from
the interference between the Bethe-Heitler and the virtual Compton processes.
Its contribution is equal to
\begin{eqnarray}
\label{dsec}
&&\Delta d\sigma^{(I)}(\rightarrow;\leftarrow)=
(dPS)^3 2 \frac{e^6}{Q^2 \hat t}
L^{(Inter)}_{\mu\nu, \alpha}{\cal I}m H_{\mu\nu, \alpha}^{(Inter)},
\end{eqnarray}
where $(dPS)^3$ is the three-particles  phase space, and
the leptonic tensor $L^{(Inter)}_{\mu\nu, \alpha}$ is given by
the following trace
\begin{eqnarray}
\label{Interlepten}
L^{(Inter)}_{\mu\nu, \alpha}=
{\rm tr}\Biggl( \gamma_{\mu} \hat k
\biggl\{
\gamma_{\nu}\frac{\hat k^{\prime} +\hat\Delta}{(k^{\prime}+\Delta)^2}
\gamma_{\alpha}+
\gamma_{\alpha}\frac{\hat k - \hat\Delta}{(k - \Delta)^2}
\gamma_{\nu}
\biggr\} \hat k^{\prime} \Biggr).
\end{eqnarray}
Finally, the hadronic tensor $H^{(Inter)}_{\mu\nu, \alpha}$ is given by
\begin{eqnarray}
\label{Interhadten}
H^{(Inter)}_{\mu\nu, \alpha}=T_{\mu\nu}^{DVCS}
(p+p^{\prime})_{\alpha}F_+(t).
\end{eqnarray}

The second term in
(\ref{difcross}), $\Delta d\sigma^{(S)}(\rightarrow;\leftarrow)$,
is related to the square of the virtual Compton amplitude and
is defined by the expression
\begin{eqnarray}
\label{dsec2}
&&\Delta d\sigma^{(S)}(\rightarrow;\leftarrow)=
(dPS)^3 2 \frac{e^6}{Q^4}
L^{(S)}_{\mu\nu}{\cal I}m H_{\mu\nu}^{(S)},
\end{eqnarray}
where
the leptonic tensor $L^{(S)}_{\mu\nu}$ is
\begin{eqnarray}
\label{Slepten}
L^{(S)}_{\mu\nu}=
{\rm tr}\Biggl( \gamma_{\mu} \gamma_5 \hat k \gamma_{\nu}
\hat k^{\prime} \Biggr),
\end{eqnarray}
and the hadronic tensor $H^{(S)}_{\mu\nu}$ is given by
\begin{eqnarray}
\label{Shadten}
H^{(S)}_{\mu\nu}=T_{\mu\alpha}^{DVCS}(T_{\nu\alpha}^{DVCS})^+.
\end{eqnarray}

Calculating the traces  in (\ref{Interlepten}) and (\ref{Slepten}),
and the imaginary parts of (\ref{Interhadten}) and (\ref{Shadten}),
which arise in
the DVCS amplitude as  discussed before, we obtain for the first term in
the difference of cross sections~\cite{Ani00}
\begin{eqnarray}
\label{dsec2I}
&&\Delta d\sigma^{(I)}(\rightarrow;\leftarrow)=
(dPS)^3
\frac{e^6F_+(t) 4\xi}{q^2 t (k-\Delta)^2 (k^{\prime}+\Delta)^2}
\varepsilon_{kk^{\prime}P \Delta}
\nonumber\\
&&\int dx \Biggl( \delta(x+\xi)-\delta(x-\xi) \Biggr)\cdot
\Biggl(
H_1(x)(( k + k^{\prime})\cdot P)
+
\Biggr.
\nonumber\\
\Biggl.
&&2 H_3(x)(k^{\prime}\cdot \Delta^T) +
\frac{2\xi}{x(P\cdot Q)}H_A(x)\biggl(
(k\cdot\Delta)(k^{\prime}\cdot P) - (k^{\prime}\cdot\Delta)(k\cdot P)
\biggr)
\Biggr),
\end{eqnarray}
and for the second term
\begin{eqnarray}
\label{dsec2S}
&&\Delta d\sigma^{(S)}(\rightarrow;\leftarrow)=
(dPS)^3 \frac{e^6}{q^4}\varepsilon_{kk^{\prime}P \Delta }
\frac{2\xi}{(P\cdot Q)}
\int dxdx^{\prime}
\nonumber\\
&&\Biggl(
\biggl[ \delta(x+\xi)-\delta(x-\xi) \biggr]
\biggl[ \frac{{\cal P}}{x^{\prime}-\xi}+
\frac{{\cal P}}{x^{\prime}+\xi} \biggr] -
\Biggr.
\nonumber\\
&&\Biggr.
\biggl[ \delta(x^{\prime}+\xi)-\delta(x^{\prime}-\xi) \biggr]
\biggl[ \frac{{\cal P}}{x-\xi}+\frac{{\cal P}}{x+\xi} \biggr]
\Biggr)\cdot
\nonumber\\
&&\Biggl(
H_1(x)H_3(x^{\prime}) - H_1(x^{\prime})H_3(x) +
\Biggr.
\nonumber\\
&&\Biggl.
\biggl[ H_1(x^{\prime})\frac{H_A(x)}{x} -
H_1(x)\frac{H_A(x^{\prime})}{x^{\prime}}
\biggr] \xi
\Biggr).
\end{eqnarray}
Let us recall the notation for (\ref{dsec2I}) and (\ref{dsec2S}):
$F_+(t)$ is the pion electromagnetic form factor, arising from
the Bethe-Heitler diagrams, $k$ and $k^{\prime}$ denote the
momenta of the initial and final electron.

A careful analysis of Eqs. (\ref{dsec2I}) and (\ref{dsec2S})
shows that the twist-3 parametrizing
functions (GPDs) $H_3$ and $H_A$
appear in (\ref{dsec2I}) as corrections to 
the twist-2
parametrizing function $H_1$, while in Eq.(\ref{dsec2S}), the
twist-3 parametrizing functions, appear in the leading terms, and 
therefore, it is important to emphasize that they give the main
contribution.

Next, we turn to the consideration of
the DVCS process off pions with unpolarized leptons.
The cross section for this case is
\begin{eqnarray}
\label{dsecun}
d\sigma_{unp}=(dPS)^3 \biggl(
|T_{BH}|^2 + |T_{VC}|^2 + (T_{BH}T_{VC}^* + T_{VC}T_{BH}^*)
\biggr).
\end{eqnarray}
The pure Bethe-Heitler contribution to the
cross section reads (see \cite{Ji97b})
\begin{eqnarray}
|T_{BH}|^2=\frac{e^6}{t^2}L_{\mu\nu}^{(BH)}H_{\mu\nu}^{(BH)},
\end{eqnarray}
where the leptonic tensor $L^{(BH)}_{\mu\nu}$ is defined by
\begin{eqnarray}
\label{BHlepten}
L^{(BH)}_{\mu\nu}=
{\rm tr}\Biggl[
\biggl\{
\hat\varepsilon^*\frac{\hat k -\hat\Delta}{(k-\Delta)^2}
\gamma_{\mu}+
\gamma_{\mu}\frac{\hat k^{\prime}+\hat\Delta}{(k^{\prime}+\Delta)^2}
\hat\varepsilon^*
\biggr\} \hat k \biggl\{
\hat\varepsilon\frac{\hat k^{\prime} +\hat\Delta}{(k^{\prime}+\Delta)^2}
\gamma_{\nu}+
\gamma_{\nu}\frac{\hat k - \hat\Delta}{(k - \Delta)^2}
\hat\varepsilon
\biggr\} \hat k^{\prime}
\Biggr],
\nonumber\\
\end{eqnarray}
and the hadronic tensor $H^{(BH)}_{\mu\nu}$ is defined as
\begin{eqnarray}
\label{BHhadten}
H^{(BH)}_{\mu\nu}=\Gamma_{\mu}\Gamma_{\nu}=(p+p^{\prime})_{\mu}
(p+p^{\prime})_{\nu}F^2_+(t) .
\end{eqnarray}

For the virtual Compton process we have
\begin{eqnarray}
|T_{VC}|^2=-\frac{e^6}{Q^4}L_{\mu\nu}^{(VC)}H_{\mu\nu}^{(VC)},
\end{eqnarray}
where the leptonic tensor is given by
\begin{eqnarray}
\label{Comlepten}
L^{(VC)}_{\mu\nu}=
{\rm tr}\left[
\gamma_{\mu} \hat k \gamma_{\nu} \hat k^{\prime}
\right],
\end{eqnarray}
and the hadronic tensor is given by
\begin{eqnarray}
\label{Comhadten}
H^{(VC)}_{\mu\nu}=T_{\mu\alpha}^{DVCS}\varepsilon_{\alpha}^*
T_{\nu\beta}^{DVCS\, +}\varepsilon_{\beta}.
\end{eqnarray}
Here, as usual, $T_{\mu\nu}^{DVCS}$ denotes the DVCS amplitude.

The contribution arising from the interference between
the Bethe-Heitler and virtual Compton processes is given by
\begin{eqnarray}
T_{VC}T^+_{BH}+T_{BH}T^+_{VC}
=-\frac{2e^6}{Q^2 t}L_{\mu\nu, \alpha}^{(Inter)}
{\cal R}eH_{\mu\nu, \alpha}^{(Inter)},
\end{eqnarray}
where the expressions for
$L^{(Inter)}_{\mu\nu, \alpha}$ and $H_{\mu\nu, \alpha}^{(Inter)}$
have been presented above.

Furthermore,  using Eqs.(\ref{difcross}) and (\ref{dsecun}) and
inserting for the parametrizing functions (GPDs) $H_1$,
$H_3$ and $H_A$ their values calculated  within the MIT bag model, as will be
presented in the next section,
we obtain, as a function of $\phi$, the values for the SSA parameter
shown  in Figs.\ref{Pic3_both}, \ref{Picdif2}, and \ref{Picdif2_v}. 
Note that we have chosen  the kinematics
for the reaction (\ref{reaction}) which can be achieved at the HERMES
experiments (see also \cite{Her, Jlab}).

\section{GPDs within
an MIT bag model scheme.}

The Generalized Parton Distributions (GPDs) are defined as the parametrizing 
functions of light-cone matrix elements of bilocal field operators
\footnote{In eq. (\ref{me}) the triplet of pion fields 
is defined by $\pi=\{\pi^0, \pi^+, \pi^- \}$ (cf. \cite{Pol99})}
\begin{equation}
\label{me}
\int \frac{d\lambda}{2\pi} e^{i\lambda x} 
\langle \pi(p^\prime) | \bar \psi (-\frac{\lambda n}{2}) 
\Gamma  \psi (\frac{\lambda n}{2}) | \pi(p) \rangle,
\end{equation}
where $x$ is the momentum fraction of the parton, $n_\mu$ a light-cone vector to
be specified later and $\Gamma$ represents Dirac matrix structures. The
functions that we need arise from the expectation values of $\gamma_\mu$ and
$\gamma_\mu\gamma_5$. With the notation of Ref.\cite{Ani00} the
parametrizing functions appearing in the SSA parameter are defined by the
following matrix elements
\begin{eqnarray}
&&\sum\limits_{a=1}^{N_f}\, e_a^2 \,
\int \frac{d\lambda}{2\pi} e^{i\lambda x}
\langle \pi(p^{\prime}) | \bar\psi_a (-\frac{\lambda
n}{2}) \gamma_\mu \psi_a (\frac{\lambda
n}{2})| \pi(p) \rangle = H_1(x)P_\mu + H_3(x)\Delta^\perp_\mu, \\
&&\sum\limits_{a=1}^{N_f}\, e_a^2 \,
\int \frac{d\lambda}{2\pi} e^{i\lambda x}
\langle \pi(p^{\prime}) | \bar \psi_a (-\frac{\lambda
n}{2}) \gamma_\mu\gamma_5 \psi_a (\frac{\lambda
n}{2}) | \pi(p) \rangle = iH_A(x)\varepsilon_{\mu\nu\rho\sigma} P^\nu n^\rho
\Delta^\sigma_\perp,
\end{eqnarray}
where 
\begin{equation}
P= \frac{p + p^\prime}{2} \, , \, \Delta=p^\prime - p \, , \, 
\Delta_\perp = \Delta - (\Delta\cdot n).
\end{equation}
All the $H-$parametrizing functions 
are functions not only of
$x$ but also of the momentum transfer $t= \Delta^2$ and therefore connect the
parton distributions and the form factors \cite{Ji97b,Rad97}.
Moreover each parametrizing functions possesses the following properties
(see for instance \cite{Rad97}):
if $x$ belongs to the interval $[\xi, 1]$ then the
$H_i-$functions for fixed flavor
are with the quark distributions; if $x$ belongs to the
interval
$[-1, -\xi]$ then the $H_i-$functions for fixed flavor
are the anti-quark distributions;
the $H_i-$functions are the difference between the quark and 
anti-quark distributions when $x$ belong to interval $]-\xi,
\xi[$, {\it i.e.}
\begin{eqnarray}
\label{repHi}
&&H_i(x,\xi)\equiv \sum\limits_{a=1}^{N_f} e_a^2 H_i^a(x,\xi)=
\nonumber\\
&&\sum\limits_{a=1}^{N_f} e_a^2 \Biggl\{
{\cal H}_i^a(x,\xi)\Theta(-\xi\leq x \leq 1) -
{\cal H}_i^{\bar a}(x,\xi)\Theta(-1 \leq x \leq \xi)
\Biggr\}.
\end{eqnarray} 
Here the ${\cal H}_i^a$ and ${\cal H}_i^{\bar a}$ related to 
the $b^+_a \, b^-_a$ and $d^+_{\bar a} \, d^-_{\bar a}$ combinations 
of creation and annihilation operators, respectively. 

We next proceed to calculate the parametrizing functions
within the MIT bag model.  
We choose 
the kinematical variables in the Breit frame which become
\begin{equation}
p^\prime_\mu = (\overline{M} , \frac{\vec{\Delta}}{2}) \; , \;
 p_\mu = (\overline{M} , -\frac{\vec{\Delta}}{2})\; , \;
\Delta_\mu = (0, \Delta_\perp, -2\xi \overline{M}).
\end{equation}
and the light-cone vector is given by
\begin{equation}
n_\mu = \frac{1}{\overline{M}} (1,0,0,-1) .
\end{equation}
>From all these  equations it is easy to obtain the expressions for the
parametrizing functions,
\begin{eqnarray}
\label{H1}
H_1(x,\xi,t) & = & \sum_{a=1}^{N_f}e_a^2 
\int \frac{d\lambda}{2\pi} \, e^{i\lambda x} \, 
\langle \pi(p^\prime) |
{\overline \psi_a} (-\frac{\lambda
n}{2}) \; \slash \hspace{-0.2cm} n \; \psi_a (\frac{\lambda n}{2}) 
| \pi(p) \rangle, \\
\label{H3}
H_3(x,\xi,t) & = &\sum_{a=1}^{N_f}e_a^2  
\frac{1}{|\Delta_\perp|^2} \int \frac{d\lambda}{2\pi}
 \, e^{i\lambda x} \, 
 \langle \pi(p^\prime) |{\overline \psi_a} (-\frac{\lambda
n}{2})\; \slash \hspace{-0.27cm} \Delta_\perp \; \psi_a (\frac{\lambda n}{2}) 
| \pi(p) \rangle, \\
\label{HA}
H_A(x,\xi,t) & = &\sum_{a=1}^{N_f}e_a^2  
\frac{1}{|\Delta_\perp|} \int \frac{d\lambda}{2\pi}
\, e^{i\lambda x} \, 
 \langle \pi(p^\prime) | \psi_a^+ (-\frac{\lambda
n}{2})\; \Sigma_y  \; \psi_a (\frac{\lambda n}{2}) | \pi(p) \rangle ,
\end{eqnarray}
where $\Sigma_y$ is the $y$ component of the spin operator.
We need to emphasize, that within the most naive version MIT bag model,
{\it i.e.} when only confinement is taken
into account and no evolution is considered, only the valence quarks 
degrees of freedom are considered. 
As a consequence, in the $\pi^+$ case the matrix elements of Eqs.(\ref{H1}),
(\ref{H3}) and (\ref{HA}) reads
\begin{eqnarray}
\label{repHi2}
&&H_i^{(\pi^+)}(x,\xi)=\frac{4}{9}{\cal H}_i^u(x,\xi)
\Theta(-\xi\leq x \leq 1)
-\frac{1}{9}{\cal H}_i^{\bar d}(x,\xi)\Theta(-1 \leq x \leq \xi)=
\nonumber\\
&&\int \frac{d\lambda}{2\pi} \, e^{i\lambda x} \, 
\langle \pi^+(p^\prime) |
{\overline \psi} (-\frac{\lambda n}{2}) \; 
\Gamma_i \; \psi (\frac{\lambda n}{2}) | \pi^+(p) \rangle , 
\end{eqnarray}
where $\Gamma_i=\{ \slash \hspace{-0.2cm} n,\  
\slash \hspace{-0.27cm} \Delta_\perp, \gamma_0\Sigma_y \}$ for
the corresponding parametrizing functions.
Similar expressions can be easily written down for the other 
terms of the pion triplet.

In order to perform the calculation we use the MIT bag model in the boosted
scheme \cite{bg},  whose virtues and
defects for this type of physics have been thouroughly discussed \cite{jms}.
The expressions for the parametrizing functions above become in this framework 
of the form
\begin{equation}
2\overline{M}  \int \frac{d\lambda}{2 \pi} e^{i\lambda x} 
\int d^3x e^{i\vec{\Delta}
\cdot \vec{r}}\langle {\overline\psi}(\vec{r} -\frac{\lambda n}{2}) \;\Gamma\; \psi(\vec{r}
+\frac{\lambda n}{2})\rangle 
\end{equation}
where now the matrix elements are calculated within the bag states normalized 
to $1$ and $\Gamma$ symbols the required Dirac operators. These field operators 
give rise to a sum over quark (anti-quark) wave functions
of the form
\begin{equation}
2\overline{M} Z \int \frac{d\lambda}{2 \pi} e^{i\lambda x} \int d^3x
e^{i\vec{\Delta}\cdot \vec{r}}
\langle {\overline \psi}_{\vec{v}}(\vec{r} -\frac{\lambda
n}{2}) \; \Gamma \;
\psi_{\vec{v}}(\vec{r}+\frac{\lambda n}{2})\rangle 
\end{equation}
where $Z$ is a normalization factor coming from the spectator particle and
the boosted wave functions are given by
\begin{equation}
\psi_{\vec{v}}(t,\vec{r}) = S(\Lambda_{\vec{v}}) \int \frac{d^3k}{(2\pi)^3}
\exp{(-i\tilde{\varepsilon}_0t - \vec{\tilde{k}} \cdot \vec{r})}
\varphi(\vec{k})
\end{equation}
All the required definitions and notation  from now on are to be found
in Refs.\cite{bg} and \cite{jms}.

After a tedious but straightforward calculation  we obtain the parametrizing
functions for each quark flavor

\begin{eqnarray}
{\cal H}_1^a(x,\xi,t) & = & \frac{4 \pi N_0 R^6 Z \overline{M}}
{\cosh{\omega}(1-(\cosh\omega -1)\frac{\Delta_z^2}{t})}
\int \frac{dk_\perp d \varphi}{(2 \pi)^3} k_\perp \nonumber \\
& &  \left[ \varphi_0(k^\prime)
 \varphi_0 (k) + \frac{\vec{k}^\prime \cdot \vec{k}}{k^\prime k}
 \varphi_1(k^\prime) \varphi_1(k) -
\frac{k_z}{k}\varphi_0(k^\prime) \varphi_1(k) - \frac{k_z^\prime}{k^\prime}
\varphi_1(k^\prime) \varphi_0(k) \right. \nonumber  \\
& &  +  (\cosh{\omega} -1) \left(\frac{1}{k v^2}
(v_z \vec{k} \cdot \vec{v} - k_z v^2)
 \varphi_0(k^\prime) \varphi_1(k) \right. \nonumber \\
& & \left. + \frac{1}{k^\prime v^2}
(v_z \vec{k^\prime} \cdot \vec{v} - k_z^\prime v^2)
 \varphi_0(k) \varphi_1(k^\prime) \right)  \nonumber  \\
& & \left. + \sinh{\omega}\left(\frac{k_z}{k v} (\vec{k^\prime} \cdot \vec{v} -
k_z^\prime v_z) - \frac{k_z^\prime}{k^\prime v}
(\vec{k} \cdot \vec{v} -
k_z v_z)\right) \varphi_1(k) \varphi_1(k^\prime) \right],\nonumber \\
& &
\end{eqnarray}

\begin{eqnarray}
{\cal H}_3^a(x,\xi,t) & = & \frac{4 \pi N_0^2 R^6 Z \overline{M}^2 }
{\cosh{\omega}(1-(\cosh\omega -1)\frac{\Delta_z^2}{t}) \Delta_x}
\int \frac{dk_\perp d \varphi}{(2 \pi)^3} k_\perp \nonumber \\
& &  \left[- \frac{k_x}{k}\varphi_0(k^\prime) \varphi_1(k)
- \frac{k_x^\prime}{k^\prime} \varphi_1(k^\prime) \varphi_0(k)
\right. \nonumber  \\
& &  +  (\cosh{\omega} -1) \left(\frac{1}{k v^2}
(v_x \vec{k} \cdot \vec{v} - k_x v^2)
 \varphi_0(k^\prime) \varphi_1(k) \right. \nonumber \\
& & \left. + \frac{1}{k^\prime v^2}
(v_x \vec{k^\prime} \cdot \vec{v} - k_x^\prime v^2)
 \varphi_0(k) \varphi_1(k^\prime) \right)  \nonumber  \\
& & \left. + \sinh{\omega}\left(\frac{k_z}{k v} (\vec{k^\prime} \cdot \vec{v} -
k_x^\prime v_x) - \frac{k_x^\prime}{k^\prime v}
(\vec{k} \cdot \vec{v} -
k_x v_x)\right) \varphi_1(k) \varphi_1(k^\prime) \right] ,\nonumber \\
& &
\end{eqnarray}
and
\begin{eqnarray}
{\cal H}_A^a(x,\xi,t) & = & \frac{4 \pi N_0 R^6 Z \overline{M} ^2}
{\cosh{\omega}(1-(\cosh\omega -1)\frac{\Delta_z^2}{t}) \Delta_x}
\int \frac{dk_\perp d \varphi}{(2 \pi)^3} k_\perp \nonumber \\
& & \left[\cosh{\omega}\frac{(\vec{k} \times \vec{k^\prime})_y}{k k^\prime}
\varphi_1(k) \varphi_1(k^\prime) \right. \nonumber \\
& & \left. + \sinh{\omega} \left(\frac{(\vec{k} \times \vec{v})_y}{k v}
 \varphi_0(k^\prime) \varphi_1(k)
 + \frac{(\vec{k^\prime} \times \vec{v})_y}{k^\prime v}
 \varphi_0(k) \varphi_1(k^\prime) \right) \right]. \nonumber \\
& &
\end{eqnarray}

The valence anti-quark corresponding functions are obtained from these 
by the following transformations
\begin{eqnarray}
{\cal H}_1^{\bar{a}}(x,\xi,t) & = & {\cal H}_1^a(-x,\xi,t), \nonumber \\
{\cal H}_3^{\bar{a}} (x,\xi,t)  & = & {\cal H}_3^a(-x,\xi,t), \nonumber \\
{\cal H}_A^{\bar{a}} (x,\xi,t)  & = &  -{\cal H}_A^a (-x,\xi,t).
\nonumber \\
& &
\end{eqnarray}

The calculation thus far suffers from a traditional problem, namely the so
called support problem, {\it i.e.} the parametrizing functions are non-vanishing
outside the physical range $[-1,1]$. For simplicity we use a generalization 
of the prescription of Ref.(\cite{traini}) given by 
\begin{equation}
h(x) \rightarrow \frac{1}{(1-|x|)^2} \; h(\frac{x}{1-|x|}),
\end{equation}
which limits the functions to the adequate interval
$[-1,1]$, but does not avoid that the quark (anti-quark) contribution 
extends into the negative (positive)
$x$ region: our partons are only quark (anti-quark) valence partons and therefore the functions
should not extend to negative (positive) $x$. However this deffect has a minor
impact on the final result. 

Using the above expressions adequately modified by the support prescription
and saturating the spin flavor degrees of freedom of
the pion wave functions we obtain the  pion parametrizing functions 
within the boosted scheme which are
shown in Figs.\ref{H1_v}, \ref{H3_v}, and \ref{HA_v}. 
In them one can see how the region around $x=0$ is
the most problematic, but does not affect  the asymmetries in an important
manner.

For the sake of completeness and complementarity we have
performed the calculation also in the unboosted Peirls-Yoccoz \cite{py} 
scheme. The latter has no support problem, but lacks 
recoil corrections. In this case we obtain the following
equations, 
\begin{eqnarray}
\label{eq63}
\tilde{\cal H}^{u}_{1}(x,\xi,t)&=&N^{2}(4\pi R^{3})^{2}
\int_{0}^{\infty}\vert k_{\perp}\vert \frac{d\vert k_{\perp}\vert}{(2\pi)^2} 
\int_{0}^{2\pi} \frac{d\varphi_{\alpha}}{(2\pi)}
\frac{\vert\phi_{1}({\bf k+\frac{\Delta}{2}})\vert^2}{\vert\phi_{2}({\bf \Delta}/2)\vert^2}
\frac{\vert (k+\frac{\Delta}{2})_{0}\vert}{\vert (1-x)\vert}\nonumber\\
&\times&\Bigg(\varphi_{0}(k)\varphi_{0}(k')
+\frac{\vec{k}·\vec{k}'}{k k'}\varphi_{1}(k)\varphi_{1}(k')\nonumber\\
&-&\frac{k_{z}-\Delta_{z}/2}{k}\varphi_{0}(k')\varphi_{1}(k)
-\frac{k'_{z}+\Delta_{z}/2}{k'}\varphi_{0}(k)\varphi_{1}(k')
\Bigg), 
\end{eqnarray}
\begin{eqnarray}
\label{eq64}
\tilde{\cal H}^{u}_{3}(x,\xi,t)&=&\frac{1}{2}N^{2}(4\pi R^{3})^{2}\bar{M}
\int_{0}^{\infty}\vert k_{\perp}\vert \frac{d\vert k_{\perp}\vert}{(2\pi)^2} 
\int_{0}^{2\pi} \frac{d\varphi_{\alpha}}{(2\pi)}
\frac{\vert\phi_{1}({\bf k+\frac{\Delta}{2}})\vert^2}{\vert\phi_{2}({\bf \Delta}/2)\vert^2}
\frac{\vert (k+\frac{\Delta}{2})_{0}\vert}{\vert (1-x)\vert}\nonumber\\
&\times&\Bigg(\frac{1}{k}\varphi_{0}(k')\varphi_{1}(k)
+\frac{1}{k'}\varphi_{0}(k)\varphi_{1}(k')
\Bigg), 
\end{eqnarray}
and 
\begin{eqnarray}
\label{eq65}
\tilde{\cal H}^{u}_{A}(x,\xi,t)&=&\frac{1}{2}N^{2}(4\pi R^{3})^{2}\bar{M}
\int_{0}^{\infty}\vert k_{\perp}\vert \frac{d\vert k_{\perp}\vert}{(2\pi)^2} 
\int_{0}^{2\pi} \frac{d\varphi_{\alpha}}{(2\pi)}
\frac{\vert\phi_{1}({\bf k+\frac{\Delta}{2}})\vert^2}{\vert\phi_{2}({\bf \Delta}/2)\vert^2}
\frac{\vert (k+\frac{\Delta}{2})_{0}\vert}{\vert (1-x)\vert}\nonumber\\
&\times&\frac{k_{z}}{k k'}\varphi_{1}(k)\varphi_{1}(k').
\end{eqnarray}
Here the normalization of the wave functions reads:
\begin{equation}
\vert\phi_{2}({\bf p})\vert ^{2}=\frac{4\pi R^{3}}{(\omega^{2}-sin^{2}\omega)^{2}}
\frac{1}{u}\int_{0}^{\omega}\frac{dv}{v}sin\big(\frac{2uv}{\omega}\big)T^{2}(v),
\end{equation}
\begin{equation}
\vert\phi_{1}({\bf p})\vert ^{2}=\frac{4\pi R^{3}}{(\omega^{2}-sin^{2}\omega)}
\int_{0}^{\omega}\frac{vdv}{\omega^{2}}\frac{sin\big(\frac{2uv}{\omega}\big)}
{\frac{2uv}{\omega}}T(v),
\end{equation}
with 
\begin{equation}
v=\frac{\vert{\bf r}\vert\omega}{2R}, \qquad u=\vert{\bf p}\vert R,
\end{equation}
and the funtion $T(v)$ given by
\begin{equation}
T(v)=\bigg[\omega-\frac{\sin^{2}\omega}{\omega}-v\bigg]\sin2v-\bigg[\frac{1}{2}
+\frac{\sin2\omega}{2\omega}\bigg]\cos2v+\frac{1}{2}+\frac{\sin2\omega}{2\omega}
-\frac{\sin^{2}\omega}{\omega^{2}}v^{2}.
\end{equation}

The results corresponding to the Peirls-Yokkoz scheme are
shown in Figs. \ref{H1_Igor}, \ref{H3_Igor}, and
\ref{HA_Igor}.

For $t\rightarrow 0$ both approaches are almost the same.
However as $t$ grows they become different, and in
particular the $H_3$ parametrizing function, which is
strongly dependent on the boost, becomes very
small in the boosting scheme. This is the reason behind the
smallness of the twist-3 contribution to the asymmetry in the
latter. 

As a final check we calculate the hadron sum rules that
 arise from $T-$invariance \cite{Pol99,Ani00},
\cite{Kiv01}, {\it i.e.}
\begin{eqnarray}
\label{sr1}
&&\int\limits_{-1}^{1} dx \Biggl\{
\frac{2}{3}{\cal H}_1^{u(d)}(x,\xi)
\Theta(-\xi\leq x \leq 1)
+\frac{1}{3}{\cal H}_1^{\bar d(\bar u)}(x,\xi)\Theta(-1 \leq x \leq \xi)
\Biggr\}=F_\pi(t)
\\
\label{sr2}
&&\int\limits_{-1}^{1} dx \Biggl\{
{\cal H}_{(3,A)}^{u(d)}(x,\xi)
\Theta(-\xi\leq x \leq 1)
-{\cal H}_{(3,A)}^{\bar u(\bar d)}(x,\xi)\Theta(-1 \leq x \leq \xi)
\Biggr\}=0
\end{eqnarray} 
We reproduce these sum rules with good precision
within our model calculations. In particular when $t\to 0$ 
we get for the pion form factor (\ref{sr1}) $0.94$ instead
of one.      

\section{Concluding Remarks}

The study of DVSC has shown that electromagnetic gauge invariance requires 
twist-3 contributions. In order to check this result experimentally we have analyzed
the scattering of linearly polarized electrons off hadrons and demonstrated that
the SSA is in principle sensitive to the twist-3 contribution. In order to be
quantitative we have been forced to calculate GPDs, in particular, certain
parametrizing functions which characterize the needed GPDs. To do so we have
performed a calculation of the required lightcone matrix elements
in the MIT bag model with both boosted and unboosted wave functions.

In Figs.\ref{H1_v}---\ref{HA_Igor} we show all the parametrizing functions. 
We have performed the calculation in two complementary
schemes. The boosting scheme, which takes proper care of the
recoil of the pion, but does not deal in an exact manner with the
center of mass problem and the Peirls Yoccoz scheme, which
contains no treatment of the recoil, but deals adequately,
for small $t$, with the center of mass problem.
Certainly the twist-2 
$H_1$ is the largest, but the twist-3 ones are non negligible and 
therefore, with
an appropriate choice of kinematics, they could be even dominant. 
However we have 
restricted our choice to the kinematics which can be achieved in HERMES. 
For the nucleon, which contains only
valence quarks in our scheme, the contribution of the 
twist-3 parametrizing functions 
will be larger.

\begin{figure}
\begin{center}
\includegraphics[width=6.0cm,height=8.0cm,angle=-90]{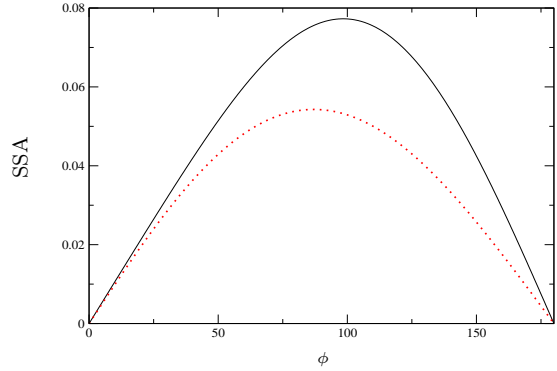}
\put(-115,-160){\tiny{$\phi$}}
\put(-230,-90){\rotatebox{90}{\scriptsize{SSA}}}
\caption{The SSA parameter as function of angle $\phi$.
The kinematic region of the calculation is defined by:
 $\hat t$=-0.1 $GeV^2$, x=0.3, $\xi$=0.3, S=22. $GeV^2$.
 Solid curve: the result within the unboosted scheme.
 Dashed curve: the resilt within the boosted scheme.}
\label{Pic3_both}
\end{center}
\end{figure}

\begin{figure}
\begin{center}
\includegraphics[width=6.0cm,height=8.0cm,angle=-90]{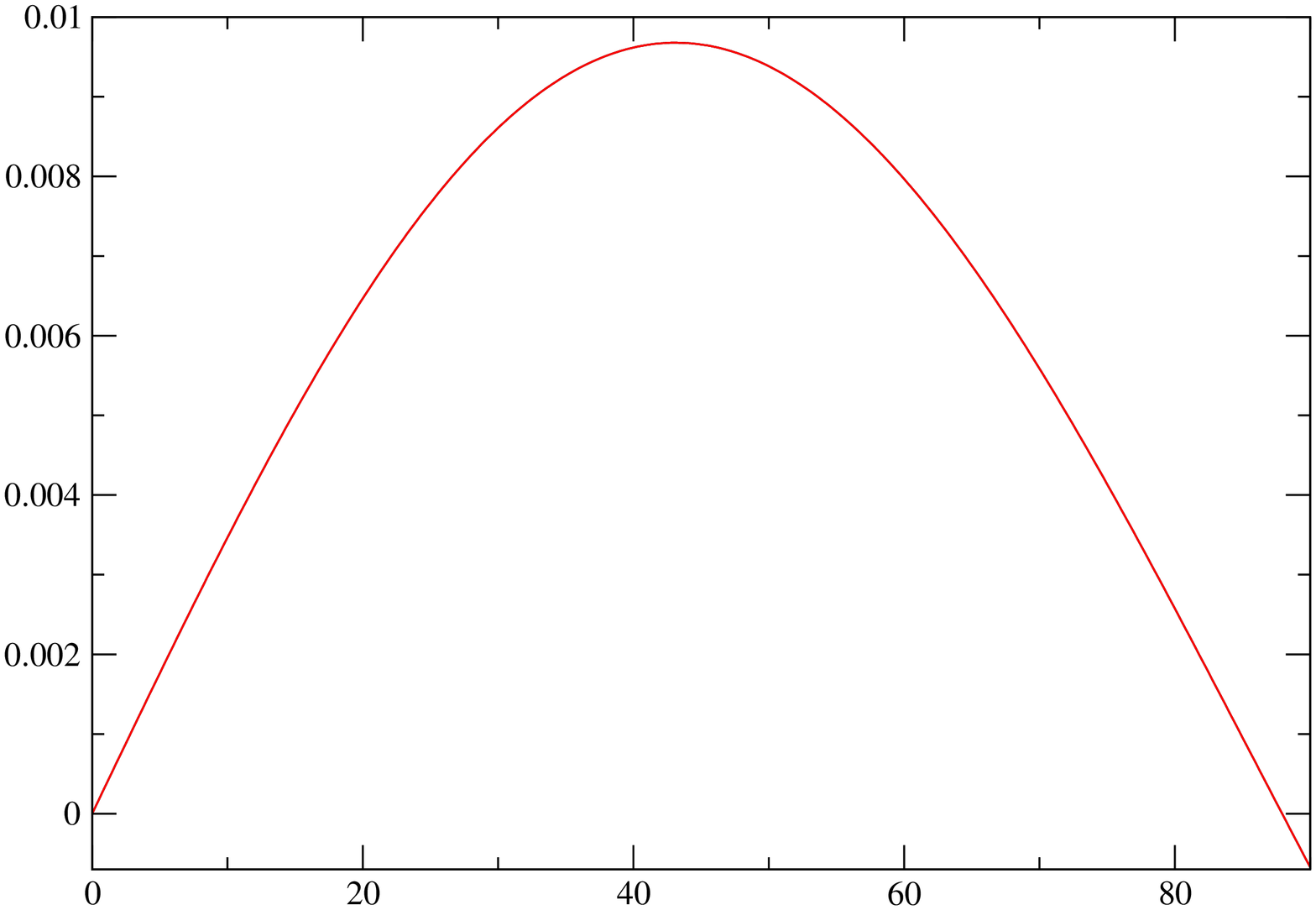}
\put(-115,-160){\tiny{$\phi$}}
\put(-230,-90){\rotatebox{90}{\scriptsize{SSA}}}
\caption{The unboosted scheme SSA parameter from  only twist-3
contribution as a function of $\phi$ for the same kinematics as in 
Fig.\ref{Pic3_both}.}
\label{Picdif2}
\end{center}
\end{figure}

\begin{figure}
\begin{center}
\vspace{-0.5cm}
\includegraphics[width=6.0cm,height=8.0cm,angle=-90]{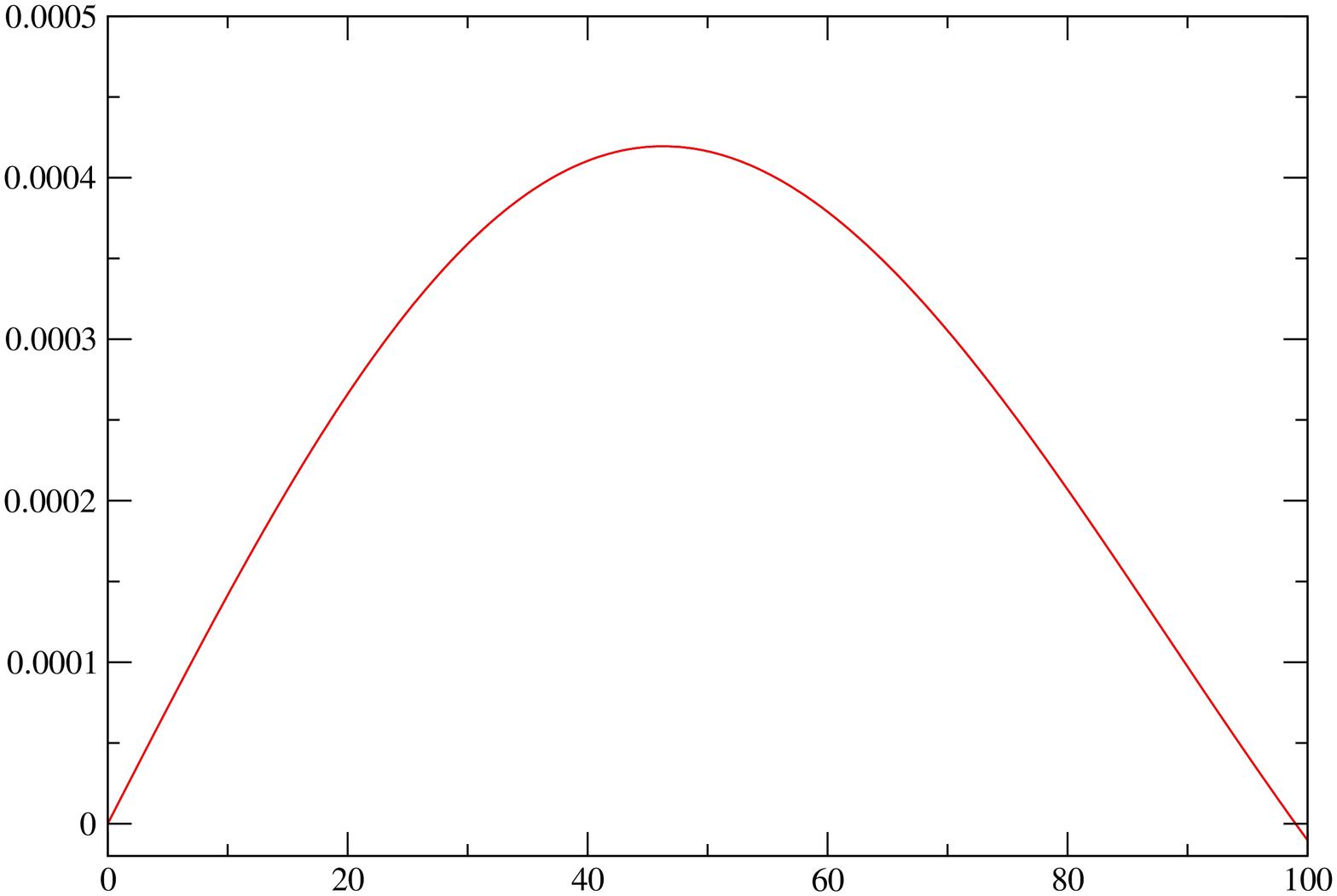}
\put(-115,-160){\tiny{$\phi$}}
\put(-230,-90){\rotatebox{90}{\scriptsize{SSA}}}
\caption{The boosted scheme SSA parameter from  only twist-3
contribution as a function of $\phi$ for the same kinematics as in 
Fig.\ref{Pic3_both}.}
\label{Picdif2_v}
\end{center}
\end{figure}

\begin{figure}
\begin{center}
\includegraphics[height=8.0cm,angle=-90]{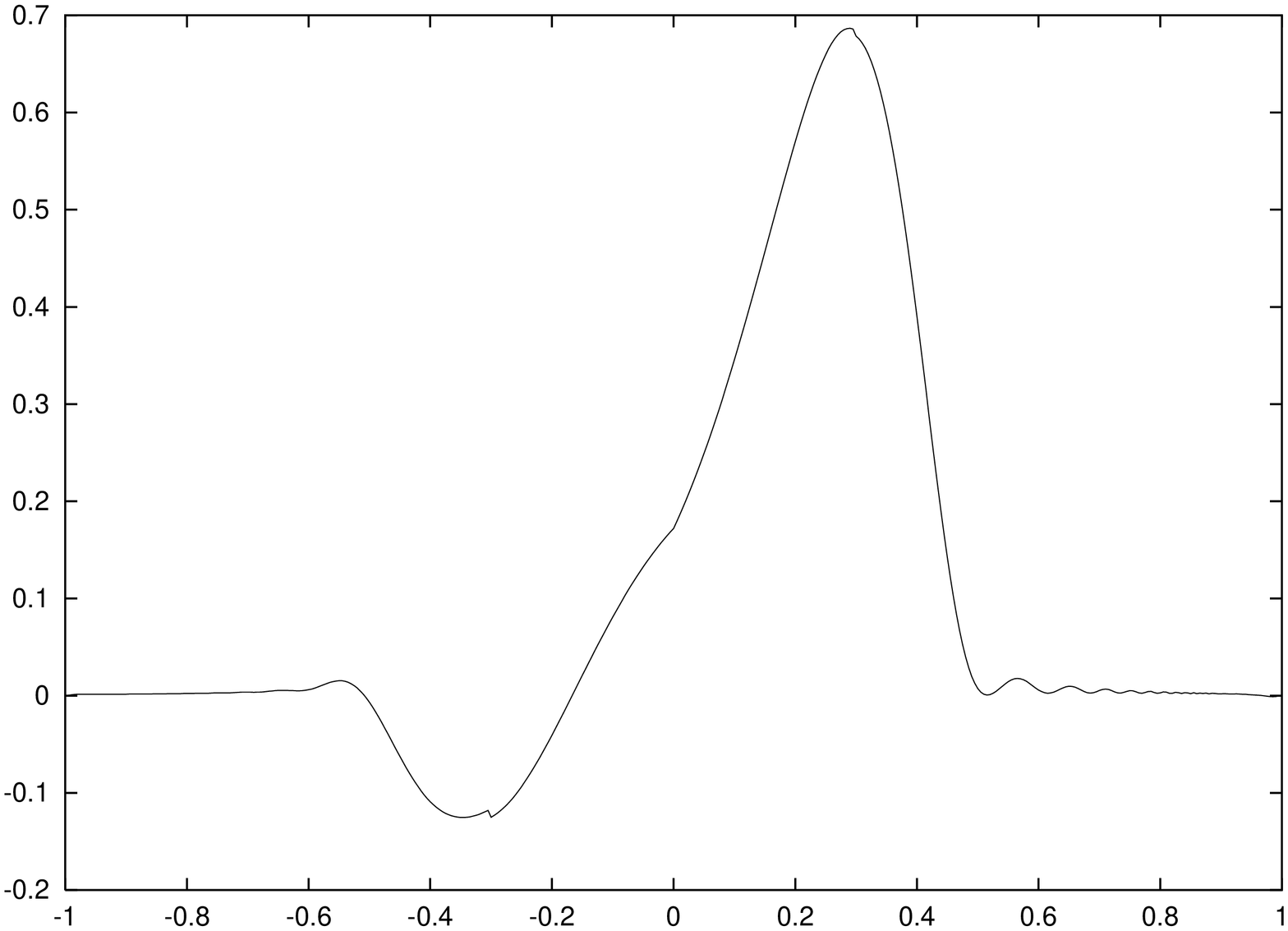}
\put(-77,-115){\tiny{$x$}}
\put(-165,-87){\rotatebox{90}{\tiny{$H_1(x,0.3),\quad
t=-0.1\, GeV^2$}}}
\vspace{0.2cm}
\caption{The boosted scheme generalized parton distribution
$H_1$ for the 
$\pi^\pm$ case
at $t=-0.1 \, GeV^2$ and $\xi=0.3$.}
\label{H1_v}
\end{center}
\end{figure}

\begin{figure}
\begin{center}
\includegraphics[height=8.0cm,angle=-90]{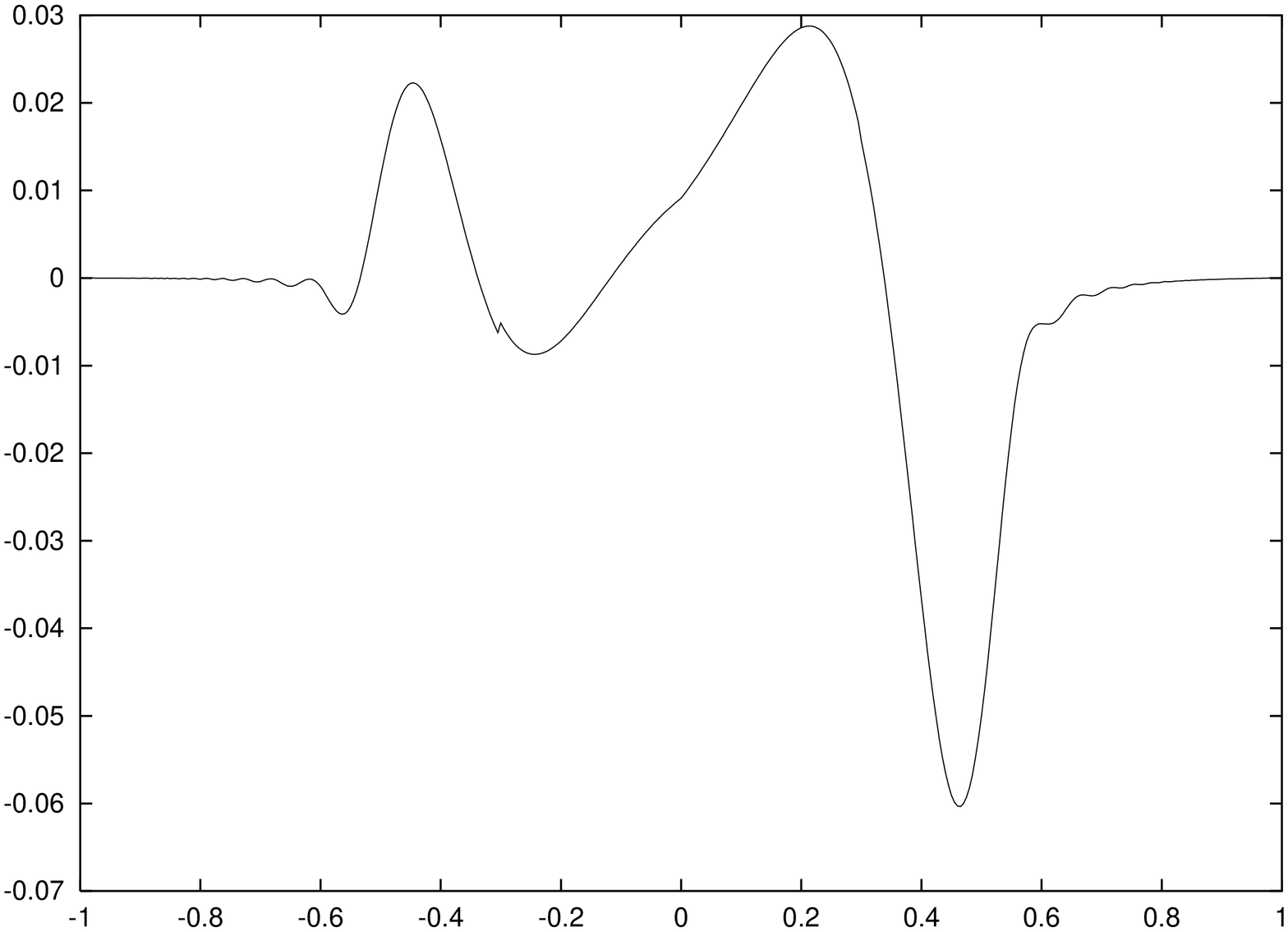}
\put(-77,-115){\tiny{$x$}}
\put(-165,-87){\rotatebox{90}{\tiny{$H_3(x,0.3),\quad
t=-0.1\, GeV^2$}}}
\vspace{0.2cm}
\caption{The boosted scheme generalized parton distribution
$H_3$ for the 
$\pi^\pm$ case and for the same kinematics as in
Fig.\ref{H1_v}.}
\label{H3_v}
\end{center}
\end{figure}

\begin{figure}
\begin{center}
\includegraphics[height=8.0cm,angle=-90]{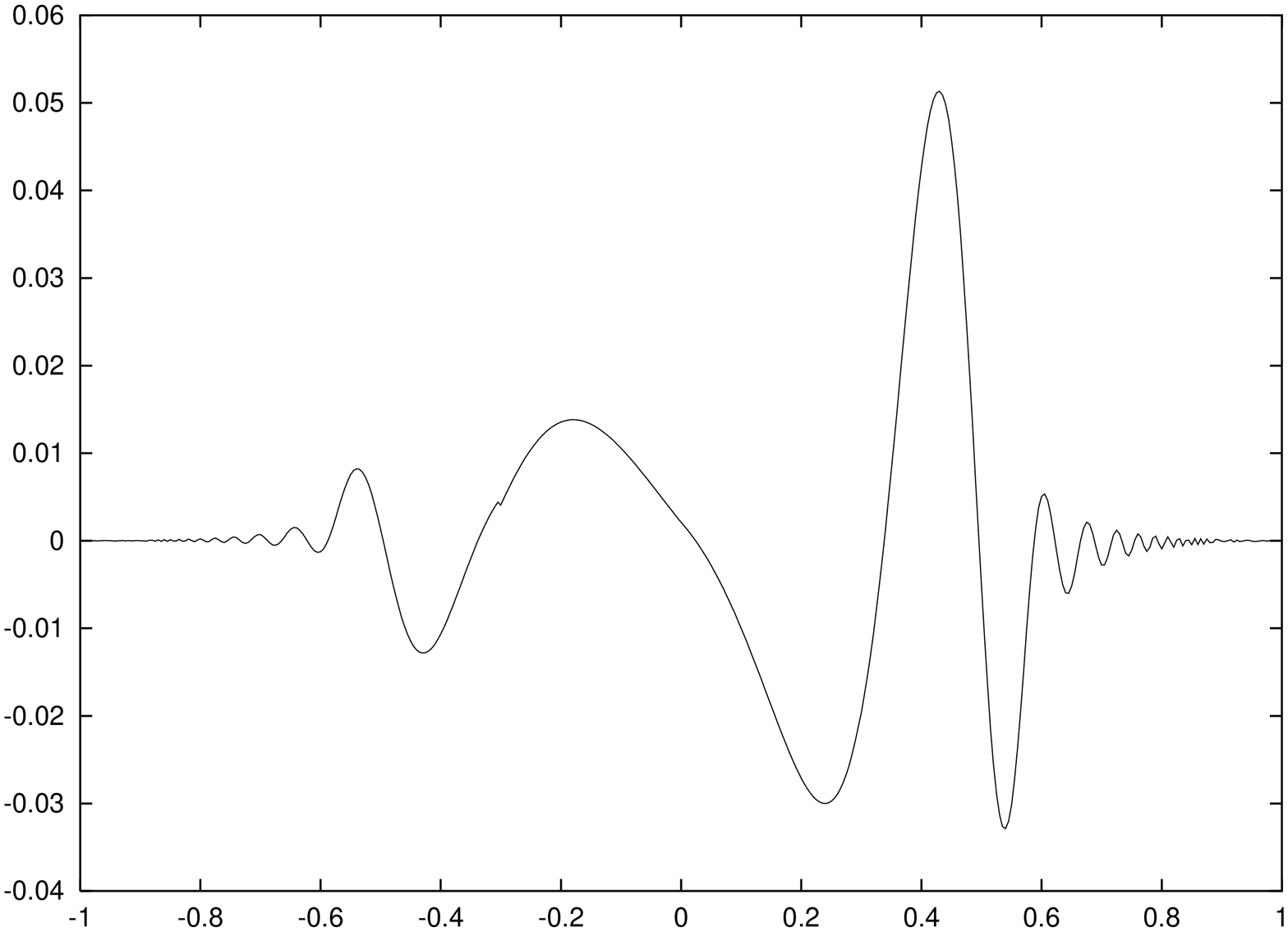}
\put(-77,-115){\tiny{$x$}}
\put(-165,-87){\rotatebox{90}{\tiny{$H_A(x,0.3),\quad
t=-0.1\, GeV^2$}}}
\vspace{0.2cm}
\caption{The boosted scheme generalized parton distribution
$H_A$ for the 
$\pi^\pm$ case
and for the same kinematics as in Fig.\ref{H1_v}.}
\label{HA_v}
\end{center}
\end{figure}

\begin{figure}
\begin{center}
\includegraphics[height=8.0cm,angle=-90]{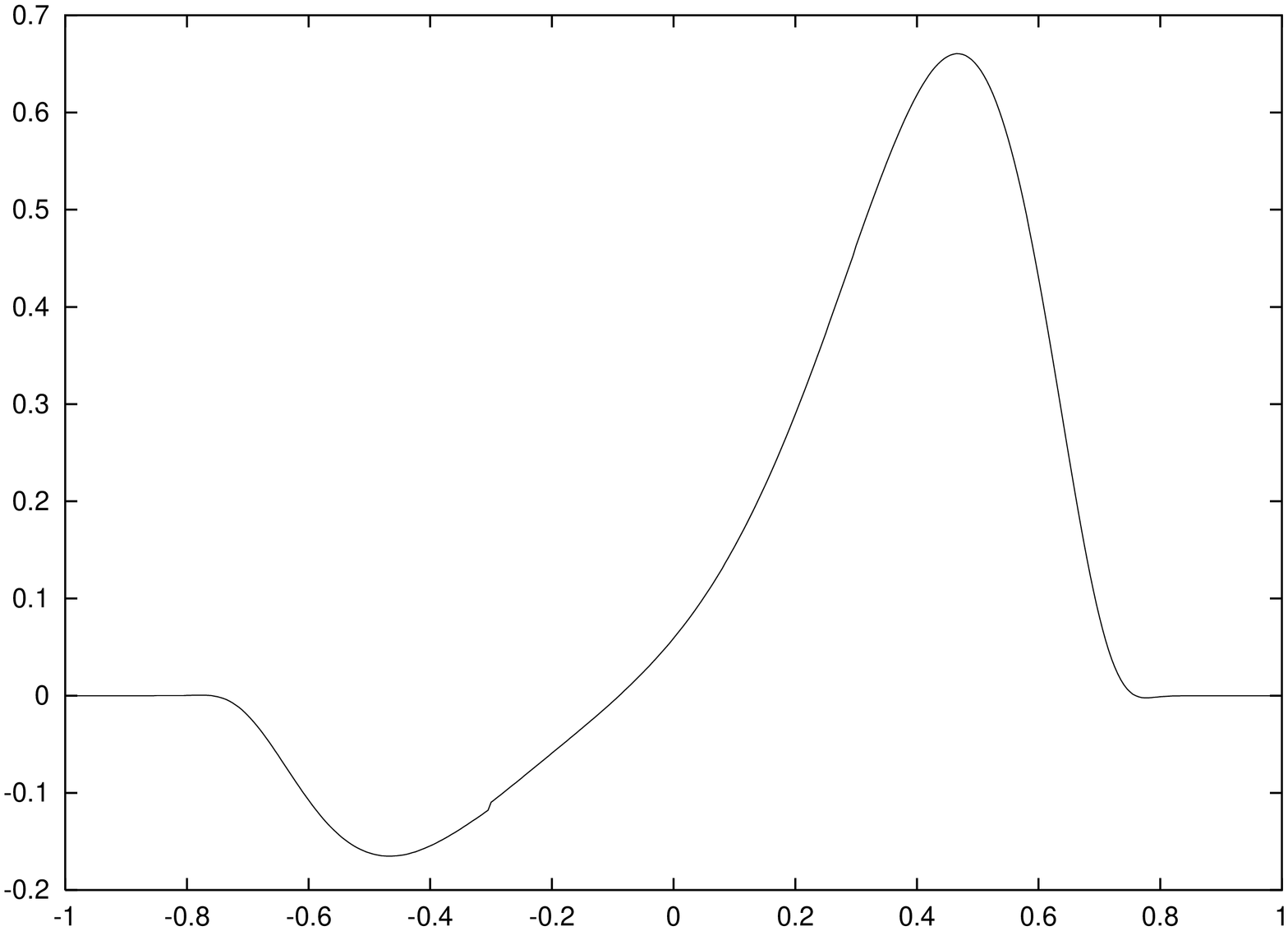}
\put(-77.5,-115){\tiny{$x$}}
\put(-165,-87){\rotatebox{90}{\tiny{$H_1(x,0.3),\quad t=-0.1\, GeV^2$}}}
\vspace{0.2cm}
\caption{The unboosted scheme generalized parton distribution $H_1$ for the 
$\pi^\pm$ case
at $t=-0.1 \, GeV^2$ and $\xi=0.3$.}
\label{H1_Igor}
\end{center}
\end{figure}

\begin{figure}
\begin{center}
\includegraphics[height=8.0cm,angle=-90]{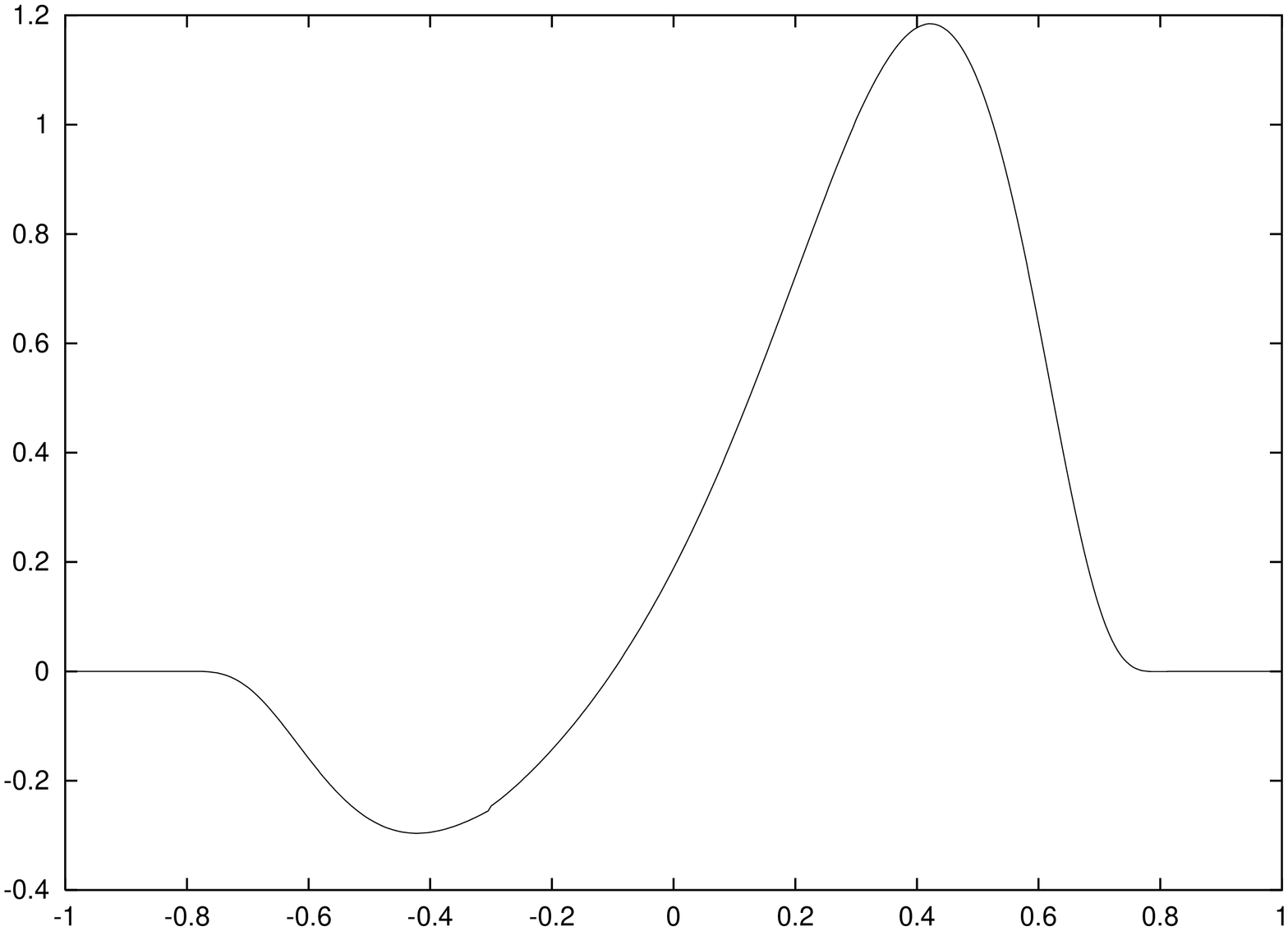}
\put(-77.5,-115){\tiny{$x$}}
\put(-165,-87){\rotatebox{90}{\tiny{$H_3(x,0.3),\quad t=-0.1\, GeV^2$}}}
\vspace{0.2cm}
\caption{The unboosted scheme generalized parton distribution $H_3$ for the 
$\pi^\pm$ case and for the same kinematics as in Fig.\ref{H1_Igor}.}
\label{H3_Igor}
\end{center}
\end{figure}

\begin{figure}
\begin{center}
\includegraphics[height=8.0cm,angle=-90]{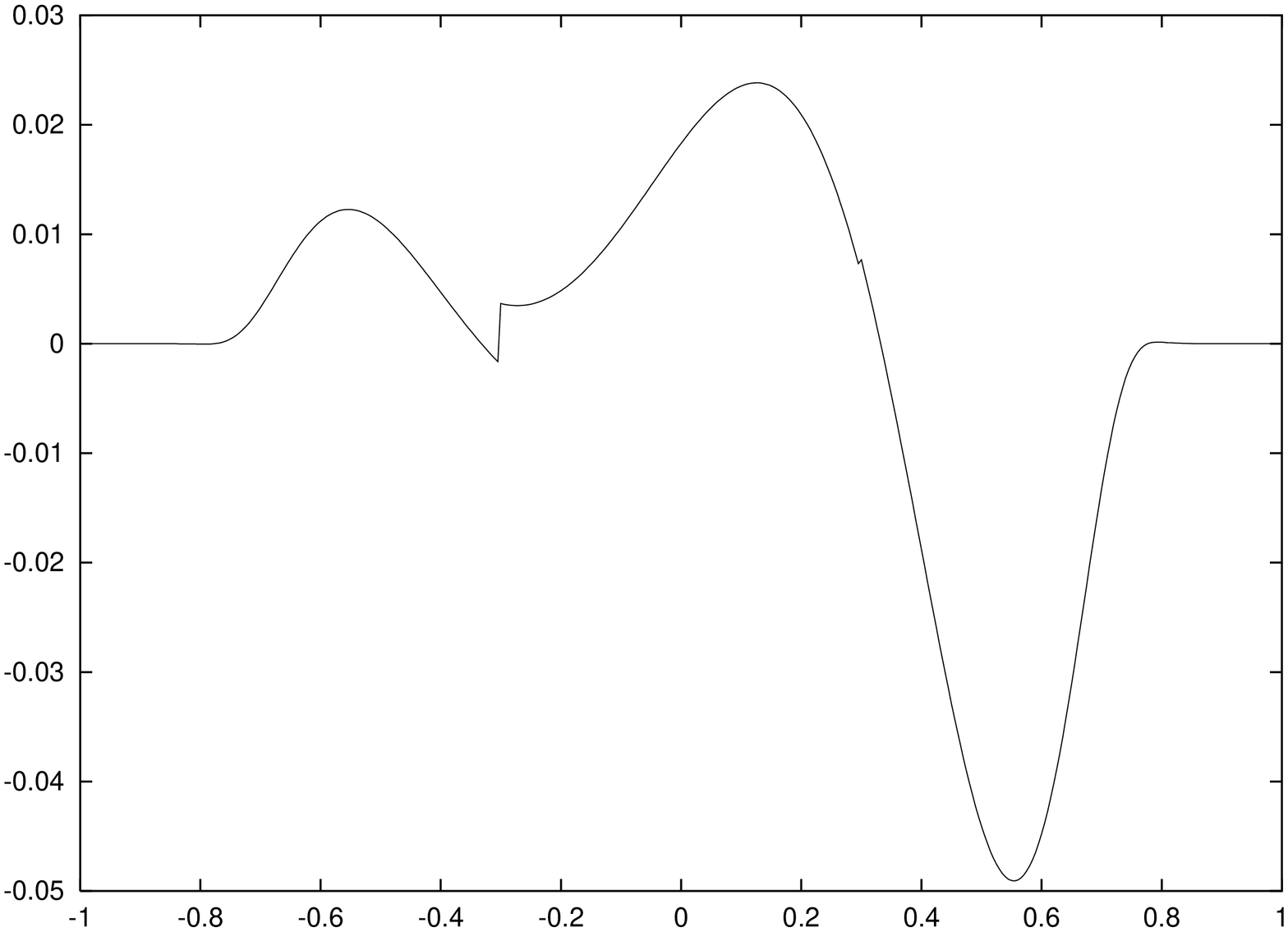}
\put(-77.5,-115){\tiny{$x$}}
\put(-165,-87){\rotatebox{90}{\tiny{$H_A(x,0.3),\quad t=-0.1\, GeV^2$}}}
\vspace{0.2cm}
\caption{The unboosted scheme generalized parton distribution $H_A$ for the 
$\pi^\pm$ case
and for the same kinematics as in the Fig.\ref{H1_Igor}.}
\label{HA_Igor}
\end{center}
\end{figure}

In Figs. \ref{Pic3_both}, \ref{Picdif2}, 
and \ref{Picdif2_v} we show different aspects of our study of the SSA.
Fig.\ref{Pic3_both} 
shows the SSA obtained by taking all contributions into account. It is
small but measurable with todays high luminosity beams and efficient detectors.
In Fig.\ref{Picdif2} and \ref{Picdif2_v}
 we depict the pure twist-3 contribution to the SSA, 
which is not small, 15 \% at the peak within the unboosted scheme 
and is much less, 1 \% at the peak within the boosted scheme. 
Thus the implementation of gauge invariance is not only theoretically relevant 
but also quantitatively rather important.

Our model for calculating the parametrizing functions lacks at present one
crucial ingredient, namely Renormalization Group Evolution. The necessary
ingredients to implement such a program to next to leading order are not
available. Results, however, might be strongly affected by evolution as has 
been the case for other structure functions \cite{scopetta}. Thus until
the actual calculation has been evolved to the energy regime under scrutiny
we will not be fully certain about the experimental relevance of this 
observable.

The future of the present calculation resides in generalizing our results to the
nucleon. For unpolarized nucleons, the only difference in the treatment, arises 
from the kinematics which is more complicated because we cannot
neglect the nucleon mass. This fact, however, should not change the qualitative
features of the present results dramatically. For the reasons mentioned above, 
we expect larger values both for the twist-2 contributions, because there are
more scatterers, as for the \mbox{twist-3} ones, because the contribution 
of the $H_A$ parametrizing function will not be reduced. The next step should 
be to proceed to the study of polarized nucleons. In this case the 
number of observables increases dramatically due to the spin structure
of the target  and a study is under way aiming to separate them in concrete
experiments and estimate their values \cite{Ani02}.

Our present work shows once again that factorization allows  the use of
models and perturbative QCD in a consistent fashion generating a predictive
scheme which is useful in guiding future experimental developments.

\section*{Acknowledgements}

This work has been supported in part by EC-RTN Network            
ESOP, Contract HPRN-CT-2000-00130, by DGESIC (Spain) under            
contracts PB97-1401-C02-01 and PB97-1227, by INTAS (Project 587, call 2000)  
and by RFFI Grant 00-02-16696. V. Vento has been supported by a travel grant 
from the University of Valencia and by KIAS (Korea) during the later stages 
of this work. 

We would like to thank A. Belitsky, A.E. Dorokhov, N.I. Kochelev, 
J. Papavassiliou, M.V. Polyakov, O.V. Teryaev,
M. Vanderhaeghen for many stimulating discussions and
M. Rho for a  critical reading of the manuscript. V. Vento
is particularly greatful to W. Melnitchouck for fruitful
correspondence regarding their calculation \cite{jms}.

\newpage

\newpage
\setlength{\unitlength}{.5mm}

\newpage

\setlength{\unitlength}{.5mm}

\end{document}